\documentclass[12pt]{article}
\usepackage[english]{babel}
\usepackage{amsmath}
\usepackage{amssymb}
\usepackage{graphicx}

\usepackage[unicode,pdfborder={0 0 1}]{hyperref}

\def\hybrid{\topmargin 0pt      \oddsidemargin 0pt
        \headheight 0pt \headsep 0pt
        \voffset=-0.5cm
        \hoffset=-0.3825in
        \textwidth 7.15in       
        \textheight 9in       
        \marginparwidth 0.0in
        \parskip 5pt plus 1pt   \jot = 1.5ex}
\catcode`\@=11
\def\marginnote#1{}

\newcount\hour
\newcount\minute
\newtoks\amorpm
\hour=\time\divide\hour by60 \minute=\time{\multiply\hour by60 \global\advance\minute by-\hour}
\edef\standardtime{{\ifnum\hour<12 \global\amorpm={am}%
        \else\global\amorpm={pm}\advance\hour by-12 \fi
        \ifnum\hour=0 \hour=12 \fi
        \number\hour:\ifnum\minute<10 0\fi\number\minute\the\amorpm}}
\edef\militarytime{\number\hour:\ifnum\minute<10 0\fi\number\minute}

\def\draftlabel#1{{\@bsphack\if@filesw {\let\thepage\relax
   \xdef\@gtempa{\write\@auxout{\string
      \newlabel{#1}{{\@currentlabel}{\thepage}}}}}\@gtempa
   \if@nobreak \ifvmode\nobreak\fi\fi\fi\@esphack}
        \gdef\@eqnlabel{#1}}
\def\@eqnlabel{}
\def\@vacuum{}
\def\draftmarginnote#1{\marginpar{\raggedright\scriptsize\tt#1}}
\def\draftlabel#1{{\@bsphack\if@filesw {\let\thepage\relax
   \xdef\@gtempa{\write\@auxout{\string
      \newlabel{#1}{{\@currentlabel}{\thepage}}}}}\@gtempa
   \if@nobreak \ifvmode\nobreak\fi\fi\fi\@esphack}
        \gdef\@eqnlabel{#1}}
\def\@eqnlabel{}
\def\@vacuum{}
\def\draftmarginnote#1{\marginpar{\raggedright\scriptsize\tt#1}}

\def\draft{\oddsidemargin -.5truein
        \def\@oddfoot{\sl preliminary draft \hfil
        \rm\thepage\hfil\sl\today\quad\militarytime}
        \let\@evenfoot\@oddfoot \overfullrule 3pt
        \let\label=\draftlabel
        \let\marginnote=\draftmarginnote
   \def\@eqnnum{(\theequation)\rlap{\kern\marginparsep\tt\@eqnlabel}%
\global\let\@eqnlabel\@vacuum}  }

\def\numberbysection{\@addtoreset{equation}{section}
        \def\theequation{\thesection.\arabic{equation}}}

\def\underline#1{\relax\ifmmode\@@underline#1\else
        $\@@underline{\hbox{#1}}$\relax\fi}

\def\titlepage{\@restonecolfalse\if@twocolumn\@restonecoltrue\onecolumn
     \else \newpage \fi \thispagestyle{empty}\c@page\z@
        \def\thefootnote{\fnsymbol{footnote}} }

\def\endtitlepage{\if@restonecol\twocolumn \else  \fi
        \def\thefootnote{\arabic{footnote}}
        \setcounter{footnote}{0}}  

\numberbysection \hybrid

\newcommand{\mat}[4]{\left(\begin{array}{cc}{#1}&{#2}\\{#3}&{#4}
\end{array}\right)}

\newcommand{\eqlb}[2]{\begin{equation} \label{#1} #2 \end{equation}}
\newcommand{\eq}[1]{\begin{equation} #1 \end{equation}}
\newcommand{\eqn}[1]{\begin{equation*} #1 \end{equation*}}

\newcommand{\eqs}[1]{$#1$}
\newcommand{\brc}[1]{\left(#1\right)}
\newcommand{\bsq}[1]{\left[#1\right]}
\newcommand{\bfi}[1]{\left\{ #1\right\}}

\newcommand{\qq}{\qquad}

\newcommand{\wt}[1]{\widetilde{#1}}
\newcommand{\matr}[2]{\begin{array}{#1}#2\end{array}}

\newcommand{\at}[2]{\genfrac{}{}{0pt}{}{#1}{#2}}

\newcommand{\rme}{\textrm{e}}
\newcommand{\rmd}{\textrm{d}}
\newcommand{\sn}{\textrm{sn}}
\newcommand{\cn}{\textrm{cn}}
\newcommand{\dn}{\textrm{dn}}

\newcommand{\la}{\lambda}
\newcommand{\vf}{\varphi}
\newcommand{\be}{\beta}

\newcommand{\ti}{\tilde}
\newcommand{\p}{\partial}

\newcommand{\tr}{\hbox{tr}}

\begin{document}

\begin{titlepage}
\setcounter{page}{1}

\title{ Three-particle Integrable Systems\\ with Elliptic Dependence on Momenta\\ and Theta Function Identities}

\vskip30mm

\author{
G. Aminov
\thanks{Institute of Theoretical and Experimental Physics,
  Moscow, Russia and Moscow Institute of Physics and Technology,
  Dolgoprudny,  Russia. E-mail: aminov@itep.ru}
 \and
 A. Mironov
\thanks{Theory Department, Lebedev Physics Institute and Institute of
 Theoretical and Experimental Physics, Moscow, Russia. E-mail: mironov@itep.ru, mironov@lpi.ac.ru}
 \and
A. Morozov
\thanks{Institute of Theoretical and Experimental Physics,
  Moscow, Russia. E-mail: morozov@itep.ru}
   \and
A. Zotov
\thanks{Institute of Theoretical and Experimental Physics and
   Steklov Mathematical Institute, RAS, Moscow, Russia;
  Moscow Institute of Physics and Technology,
  Dolgoprudny, Russia. E-mail: zotov@itep.ru}
}

\date{}

\maketitle

\vspace{-65mm} \centerline{\hfill ITEP-TH-21/13} \vspace{65mm}
\vspace{-65mm} \centerline{\hfill FIAN/TD-11/13} \vspace{65mm}

\begin{abstract}
We claim that some non-trivial theta-function  identities at higher genus can stand behind the Poisson commutativity of the
Hamiltonians of elliptic integrable systems, which were introduced in \cite{BMMM'2000,MM} and are made from the
theta-functions on Jacobians of the Seiberg-Witten curves. For the case of three-particle systems the genus-2 identities are
found and presented in the paper. The connection with the Macdonald identities is established. The genus-2 theta-function
identities provide the direct way to construct the Poisson structure in terms of the coordinates on the Jacobian of the
spectral curve and the elements of its period matrix. The Lax representations for the two-particle systems are also obtained.
\end{abstract}

\end{titlepage}

\small{

\section{Introduction} \label{sec1}
We suggest a new approach  to obtain the multi-particle  generalization of p,q-dual integrable systems constructed in
\cite{BMMM'2000,BMMM'1999}. The list of these systems includes duals to elliptic Calogero \cite{Calogero} and elliptic
Ruijsenaars models \cite{Ruj'88}. The most interesting are the double-elliptic integrable systems
\cite{BMMM'2000,MM,MM2,BGOR}, where both coordinates and momenta take values on elliptic curves. From the point of view of
the low-energy effective actions of Yang-Mills theories \cite{SW'94/1,SW'94/2} the integrable systems under consideration are
associated with (compactified) six-dimensional SUSY gauge theories \cite{W'96,GMMM'arx96,D'97,HW'97,G'97,W'97,MMM'98,GKM'98,Gai'12} and are strongly
involved in modern discussion of the Seiberg-Witten theory
\cite{GKMMM'95,DW'96,MM'2010,NS'09,NRS'11,Egor'2011,MMZZ'2013,MMRZZ'2013,MMRZZ34}.

The main claim of our paper is the existence  of some  new
theta-function identities which stand behind the Poisson
commutativity of the Hamiltonians considered in \cite{BMMM'2000,MM}.
We present such relations for the case of three-particle systems.
Unfortunately, it is not clear, how to prove them even in this
situation, but these relations seem to be rather interesting by
themselves - they can be considered as a genus two generalization of
the Macdonald identities \cite{Mac'72}. Also, these theta-function
identities provide a direct way to construct the Poisson structure
in terms of the coordinates on the Jacobian of the spectral curve
and the elements of its period matrix. In the case of three-particle
systems we reduce the problem of constructing the Poisson brackets
to the problem of solving the system of three partial differential
equations with respect to the pair of unknown functions.

In section \ref{sec:SWHam} we discuss the Hamiltonians  introduced
in \cite{BMMM'2000,MM}. The whole construction is restricted to the
Sieberg-Witten families of the spectral curves and the Poisson
commutativity of the Hamiltonians is related to the Sieberg-Witten
prepotential. In \cite{MM} this commutativity was checked in the
first non-trivial example in the first several orders of
$\Lambda/a$-expansion -- and this check was relied severely on the
known shape of the Seiberg-Witten prepotential. In section
\ref{sec:PS} we propose another approach, that deals with arbitrary
Riemann surfaces instead of the Sieberg-Witten curves. Then the
Poisson commutativity of the Hamiltonians is due to some new
theta-function identities, which seem to be true for an arbitrary
Riemann surface. In sections \ref{sec:theta2} and \ref{sec:Mac} we
present the theta-function identities of genus two for the case of
three-particle systems and reveal some interesting mathematical
structures behind them. In section \ref{sec:Ruj} the connection with
the Ruijsenaars models is established by means of some specific
trigonometric limits. In section \ref{sec:Lax} we go back to the
case of two-particle systems and construct the Lax representation
using the original Ruijsenaars' idea \cite{Ruijs}.

\section{Hamiltonians and Seiberg-Witten data}
\label{sec:SWHam} The Hamiltonians of the systems  dual to the elliptic  Calogero and Ruijsenaars models with an elliptic
dependence on momenta as well as the self-dual double elliptic systems were introduced in \cite{BMMM'2000,MM}. In the case of
\eqs{N=3} particles in the center of a mass frame one has two Hamiltonians of the form
 \eqlb{eq:thetaHam}{H_1\brc{\textbf{z}\,|\,\Omega}
=\dfrac{\theta^{(2)}_{11}(\textbf{z}|\Omega)} {\theta^{(2)}(\textbf{z}|\Omega)},\qq H_2\brc{\textbf{z}\,|\,\Omega}
=\dfrac{\theta^{(2)}_{22}(\textbf{z}|\Omega)}{ \theta^{(2)}(\textbf{z}|\Omega)},}
 where we use the following notations for
the Riemann theta function with symmetric \eqs{2\times2} period matrix $\Omega$:
 \eq{\theta^{(2)}(\textbf{z}|\Omega)=
\sum_{\textbf{n}\in\mathbb{Z}^2} \rme\brc{\dfrac12\textbf{n}^t\,\Omega\,\textbf{n}+ \textbf{n}\cdot\textbf{z}},\qq
\textbf{z}^t=(z_1,z_2),\qq \rme(x)\equiv\exp\brc{2\pi\imath x},}
 \eq{\theta^{(2)}_{\textbf{c}}\equiv\theta^{(2)}\bsq{\at{\textbf{0}}{\textbf{c}/3}}=
\sum_{\textbf{n}\in\mathbb{Z}^2}\rme\brc{\dfrac12\textbf{n}^t\Omega\textbf{n}+
\textbf{n}\cdot\brc{\textbf{z}+\frac{\textbf{c}}{3}}}\,,\ \ \textbf{c}\in\mathbb{Z}_3^2\,.} The role of the non-commutative
coordinates and momenta  in these Hamiltonians is played by the elements of the period matrix $\Omega$ and the coordinates on
the Jacobian $\textbf{z}$. The hypothesis of \cite{BMMM'2000} is that the Hamiltonians are Poisson commuting with respect to
the Seiberg-Witten symplectic structure
 \eq{\omega^{SW}=\sum_{i=1}^{N}\rmd\hat p_i\wedge\rmd a_i,}
where \eqs{\hat p_i=z_i+\frac1N\sum_{j=1}^{N}\hat p_j}, \eqs{z_3\equiv-z_1-z_2} and \eqs{\sum_{i=1}^{N}a_i=0}, while
the \eqs{N\brc{N-1}/2=3} elements of the period matrix $\Omega$ are not arbitrary, but functions of just \eqs{N-1=2} flat moduli:
\eqlb{om}{\Omega=\Omega\brc{\textbf{a}}\,.} Following this idea one can assume the Poisson brackets in terms of the
coordinates \eqs{\brc{z_1,z_2}} and the elements \eqs{\Omega_{ij}} to be of the form
 \eqlb{eq:PS0}{\bfi{z_i,\Omega_{jk}}=P_{ijk},\quad
\bfi{z_i,z_j}=0,\quad \bfi{\Omega_{ij},\Omega_{kl}}=0\,.}
Notice that the elements \eqs{P_{ijk}} are again functions of the flat moduli only. Thus, we can consider them to be the
functions of the period matrix $\Omega$ as well:
 \eqlb{eq:PS1}{P_{ijk}=P_{ijk}\brc{\Omega}\,.}
Taking into account the connection of the period matrix $\Omega$
with the Seiberg-Witten prepotential
 \eq{\Omega_{ij}=\frac{\partial^2\mathcal{F}}{\partial a_i\partial a_j},\qquad i,j\in\bfi{1,2}}
 and the assumption \cite{BMMM'2000}
that the prepotential is Poisson-commuting with the total momentum of the system, we get the following relation:
  \eqlb{eq:Fijk}{P_{ijk}=\frac{\partial^3\mathcal{F}}{\partial a_i\partial a_j\partial a_k},\qquad i,j,k\in\bfi{1,2}.}
In the Seiberg-Witten theory these third derivatives are usually given by the residue formulas and can be expressed through
theta-constants, see, for example, \cite{BMMM'2007}.

\section{Poisson brackets}
\label{sec:PS} In this section we forget about  the Seiberg-Witten prepotential and investigate, what can be achieved with
the help of (\ref{eq:PS0}) and (\ref{eq:PS1}) only. So, we consider \eqs{\Omega} as an arbitrary period matrix of genus-2
Riemann surface.

The elements of the Poisson  structure (\ref{eq:PS0}) \eqs{P_{ijk}} are unknown. To derive them as functions of $\Omega$, we
use two necessary conditions:
\begin{enumerate}
\item{the Poisson commutativity of Hamiltonians (\ref{eq:thetaHam}),}
\item{the Jacobi identity.}
\end{enumerate}
To impose the first condition we formulate the Poisson bracket
between the Hamiltonians (\ref{eq:thetaHam}) in terms of the
coordinates $\textbf{z}$ and the elements of the period matrix:
 \eq{\bfi{H_1,H_2}=\sum_{i=1}^{2}\sum_{j\leqslant k}
P_{ijk}\brc{\dfrac{\partial H_1}{\partial z_{i}}\dfrac{\partial H_2}{\partial\Omega_{jk}}-\dfrac{\partial H_2}{\partial
z_{i}}\dfrac{\partial H_1}{\partial\Omega_{jk}}}.} Then the strategy of finding $P_{ijk}$ is the following. The Poisson
commutativity of the Hamiltonians \eqs{\{H_1,H_2\}=0} means that there are some relations on genus two theta functions with
coefficients depending only on the elements of the period matrix. Hence, to satisfy the first condition we search for the
following relations on genus two theta functions:
 \eqlb{eq:HamRel}{\sum_{i=1}^{2}\sum_{j\leqslant k} P^{\alpha}_{ijk}\brc{\dfrac{\partial H_1}{\partial z_{i}}\dfrac{\partial
H_2}{\partial\Omega_{jk}}-\dfrac{\partial H_2}{\partial z_{i}}\dfrac{\partial H_1}{\partial\Omega_{jk}}}=0\,, \ \forall\,
{\bf z},\Omega\,,
}
 where index \eqs{\alpha} enumerates the relations and elements
\eqs{P^{\alpha}_{ijk}} are some new functions of \eqs{\Omega} which can be used for constructing  \eqs{P_{ijk}} in the form
 \eqlb{F}{\bfi{z_i,\Omega_{jk}}=P_{ijk}=\sum_{\alpha}F_{\alpha}P^{\alpha}_{ijk}\,,\ F_{\alpha}=F_{\alpha}\brc{\Omega}\,.}
As we mentioned earlier, it is natural to assume that the coefficients \eqs{P^{\alpha}_{ijk}} are some theta-constants. Thus,
the relations (\ref{eq:HamRel}) are made out of the theta functions, theta-constants and their derivatives. Also, the left
hand side of (\ref{eq:HamRel}) should vanish for arbitrary values of ${\bf z}$ and $\Omega$, implying that our relations are
actually some theta-function identities. It is natural to wonder, if any identities of such form exist without a direct
reference to the Seiberg-Witten theory.

Indeed, we found two relations on genus-2 theta  functions (\ref{eq:thetaHam}) and genus-2 theta constants
\eqs{P^{\alpha}_{ijk},\,\alpha=1,2}, which satisfy all the symmetry conditions arising from the previous section. The
relations are linearly independent, but the second relation can be obtained from the first one by permutating the diagonal
elements of the matrix \eqs{\Omega} and the coordinates \eqs{\brc{z_1,z_2}}. Unfortunately, it is not clear for us how to
prove these relations in general. But even the first simple analysis shows some interesting mathematical structures behind
them. We will present some details on these structures in the next two sections.

Once the relations on genus-2 theta functions with coefficients \eqs{P^{\alpha}_{ijk}} are found, we impose the second
condition (the Jacobi identity) in order to find $F_\alpha(\Omega)$ (\ref{F}). It takes the form
 \eqlb{eq:detPS0}{ \sum_{m\leqslant n}\brc{F_{1}P^{1}_{1mn}+F_{2}P^{2}_{1mn}}
\partial_{\Omega_{mn}} \brc{F_{1}P^{1}_{2ij}+F_{2}P^{2}_{2ij}}= \sum_{m\leqslant
n}\brc{F_{1}P^{1}_{2mn}+F_{2}P^{2}_{2mn}} \partial_{\Omega_{mn}} \brc{F_{1}P^{1}_{1ij}+F_{2}P^{2}_{1ij}},} where
\eqs{i\leqslant j}, \eqs{i,j\in\bfi{1,2}.}
Thus, we reduce the problem of constructing  the Poisson brackets to the problem of solving the system of three partial
differential equations with respect to the unknown functions \eqs{F_{\alpha},\,\alpha=1,2}. Of course, it is a challenging
issue to find the set of all solutions of the system (\ref{eq:detPS0}) and to understand whether the systems under
consideration are unique or not. We leave this problem for future investigation.

\section{Theta-function identities of genus 2}
\label{sec:theta2} In this section we discuss the  identities (\ref{eq:HamRel}) on the genus two Riemann theta functions and
present the elements \eqs{P^{\alpha}_{ijk},\,\alpha=1,2} as a solution of the system of linear equations. To describe the
first relation from (\ref{eq:HamRel})  with coefficients \eqs{P^{1}_{ijk}} we use the symmetry conditions generated by
(\ref{eq:Fijk})
 \eq{P^1_{112}=P^1_{211}=P^1_{121},\quad P^1_{212}=P^1_{122}=P^1_{221}} and additionally assume that
 \eq{P^1_{222}=0\,.}
These conditions leaves us with 3 independent unknown components of $P^1$: $P^1_{111}$, $P^1_{112}$ and $P^1_{212}$.
 Then substituting some particular values of ${\textbf{z}}$ in (\ref{eq:HamRel}), we get the system of
linear equations on elements \eqs{P^{1}_{ijk}}. Therefore, it is enough to consider two different values of ${\bf z}$:
 \eq{\textbf{z}_1=\brc{0,0},\qq \textbf{z}_2=\brc{0,\frac13},}
which will give the expressions of \eqs{P^{1}_{ijk}} in terms
of the derivatives of the following functions at the point \eqs{(0,0)}:
 \eq{H=H_1=\dfrac{\theta^{(2)}_{11}(\textbf{z}|\Omega)}
{\theta^{(2)}(\textbf{z}|\Omega)}= \dfrac{\theta^{(2)}_{22}(-\textbf{z}|\Omega)} {\theta^{(2)}(-\textbf{z}|\Omega)},\quad
F=H_1\brc{\textbf{z}+\textbf{z}_2}= \dfrac{\theta^{(2)}_{12}(\textbf{z}|\Omega) }{\theta^{(2)}_{01}(\textbf{z}|\Omega)},\quad
G=H_2\brc{\textbf{z}+\textbf{z}_2}= \dfrac{\theta^{(2)}_{20}(\textbf{z}|\Omega)} {\theta^{(2)}_{01}(\textbf{z}|\Omega)}.}
Solution of the above-described system of two linear equations with fixed arbitrary factor can be presented in the following way:
 \eqlb{PP1}{P^1_{111}=a_4 b_3 -a_3 b_4,\quad P^1_{112}=P^1_{211}=a_3 b_1 -a_1 b_3, \quad P^1_{212}=P^1_{122}=a_1 b_4 -a_4
b_1,\quad P^1_{222}=0,} where
 \eq{a_1=\left.\brc{\frac{\partial H}{\partial z_1}\frac{\partial H}{\partial\Omega_{11}}
}\right|_{\textbf{z}=0},\quad a_3=\left.\brc{\frac{\partial H}{\partial z_1}\frac{\partial
H}{\partial\Omega_{22}}+\frac{\partial H}{\partial z_2}\frac{\partial H}{\partial\Omega_{12}}}\right|_{\textbf{z}=0},\quad
a_4=\left.\brc{\frac{\partial H}{\partial z_2}\frac{\partial H}{\partial\Omega_{11}}+\frac{\partial H}{\partial
z_1}\frac{\partial H}{\partial\Omega_{12}}}\right|_{\textbf{z}=0},}
 \eq{b_1=\left.\brc{\frac{\partial F}{\partial
z_1}\frac{\partial G}{\partial\Omega_{11}} -\frac{\partial G}{\partial z_1}\frac{\partial F}{\partial\Omega_{11}}}
\right|_{\textbf{z}=0},\quad b_3=\left.\brc{\frac{\partial F}{\partial z_1}\frac{\partial G}{\partial\Omega_{22}}
-\frac{\partial G}{\partial z_1}\frac{\partial F}{\partial\Omega_{22}}+ \frac{\partial F}{\partial z_2}\frac{\partial
G}{\partial\Omega_{12}} -\frac{\partial G}{\partial z_2}\frac{\partial F}{\partial\Omega_{12}}} \right|_{\textbf{z}=0},}
\eq{b_4=\left.\brc{\frac{\partial F}{\partial z_2}\frac{\partial G}{\partial\Omega_{11}} -\frac{\partial G}{\partial
z_2}\frac{\partial F}{\partial\Omega_{11}}+ \frac{\partial F}{\partial z_1}\frac{\partial G}{\partial\Omega_{12}}
-\frac{\partial G}{\partial z_1}\frac{\partial F}{\partial\Omega_{12}}} \right|_{\textbf{z}=0}.}

The second relation with coefficients \eqs{P^{2}_{ijk}} can be
obtained from the first one by permutating the diagonal elements of
matrix \eqs{\Omega} and the coordinates \eqs{\brc{z_1,z_2}}. Taking
into account that the Hamiltonians (\ref{eq:thetaHam}) are invariant
under this permutation, we get the following answer for the elements
\eqs{P^{2}_{ijk}}:
 \eq{P^2_{222}=a_3 b_4 -a_4 b_3,\quad P^2_{112}=P^2_{211}=a_2 b_3 -a_3 b_2, \quad P^2_{212}=P^2_{122}=a_4 b_2 -a_2
b_4,\quad P^2_{111}=0,} where
 \eqlb{PP5}{a_2=\left.\brc{\frac{\partial H}{\partial z_2}\frac{\partial H}{\partial\Omega_{22}}
}\right|_{\textbf{z}=0},\qq b_2=\left.\brc{\frac{\partial F}{\partial z_2}\frac{\partial G}{\partial\Omega_{22}}
-\frac{\partial G}{\partial z_2}\frac{\partial F}{\partial\Omega_{22}}} \right|_{\textbf{z}=0}.}

{\bf Finally, we obtain  the identities}  (\ref{eq:HamRel}) {\bf
with the coefficients} $P^\alpha_{ijk}$ (\ref{PP1})-(\ref{PP5})}.
{\bf We emphasize that all the elements in the formula}
(\ref{eq:HamRel}) {\bf are expressed through the genus-2 theta
functions, therefore} (\ref{eq:HamRel}) {\bf is actually a
theta-function identity} (we do not present it explicitly just to
save the space -- it is somewhat lengthy). {\bf Despite $p-q$
duality requires it to be true only for the special $2$-parametric
family of period matrices $\Omega\brc{\textbf{a}}$, we claim that it
holds on entire $3$-dimensional space of period matrices $\Omega$
and for arbitrary $\textbf{z}$.} To confirm this statement we
checked it in $14$ orders of expansion in
$Q_1\equiv\rme\brc{\Omega_{11}/2}$ and
$Q_2\equiv\rme\brc{\Omega_{22}/2}$, what leaves us little doubt that
they are correct. However, a general proof is still an open problem.

As an additional evidence in support of these identities  we consider their reduction to the genus-one theta-constants -- one
would naturally expect some trivial well-known identity to occur. What happens -- it is indeed a known, but rather advanced
and interesting Macdonald identity from \cite{Mac'72}.

\section{The Macdonald identities}
\label{sec:Mac} It turns out that the relations (\ref{eq:HamRel}) can  be considered as the genus two generalizations of the
Macdonald identities \cite{Mac'72}. To show this connection we go backwards: degenerate the initial spectral curve into two
different elliptic curves by taking the limit \eqs{\Omega_{12}\rightarrow0}. Then consider the trigonometric limit
\eqs{\textrm{Im}\,\Omega_{22}\rightarrow+\infty}. Applying this procedure to the first relation with coefficients
\eqs{P^{1}_{ijk}}, we obtain the Fourier expansion in coordinate \eqs{z_2}. The first term of this expansion defines the
relation for the genus one theta functions given on an elliptic curve with modulus \eqs{\tau=\Omega_{11}}. This relation can
be written in terms of the elliptic functions $I_1$ and $I_2$ as
 \eqlb{eq:consEl}{I_1+I_2=0,}
\eq{I_1=3\frac{\theta_{00}(0)\theta_{0\frac13}(0)\theta'_{0\frac13}(0)}
{\theta_{00}(z_1)^3}\left(\theta_{00}(z_1)\theta_{0\frac23}(z_1)\theta''_{0\frac13}(z_1)-
\theta_{00}(z_1)\theta_{0\frac13}(z_1)\theta''_{0\frac23}(z_1)-2\theta'_{00}(z_1)
\theta'_{0\frac13}(z_1)\theta_{0\frac23}(z_1)+\right.}
 $$\left.+2\theta'_{00}(z_1)\theta_{0\frac13}(z_1)\theta'_{0\frac23}(z_1)\right)\,,$$
 \eq{I_2=\frac{1}{\theta_{00}(z_1)^3}\brc{\theta_{00}(z_1)\theta'_{0\frac13}(z_1)\theta_{0\frac23}(z_1)+
\theta_{00}(z_1)\theta_{0\frac13}(z_1)\theta'_{0\frac23}(z_1)-2\theta'_{00}(z_1)\theta_{0\frac13}(z_1)
\theta_{0\frac23}(z_1)}\times}
 $$\times\brc{\theta_{00}(0)\theta_{0\frac13}(0)\theta''_{0\frac13}(0)-
\theta_{0\frac13}(0)^2\theta''_{00}(0)+2\theta_{00}(0)\brc{\theta'_{0\frac13}(0)}^2}\,,$$ where the dependence of the theta
functions on modulus $\tau$ is meant.
 To simplify the relation (\ref{eq:consEl}) we change the summation variables in theta series
 (the Koizumi formula \cite{Koi'76}), which
gives the equation
$$\sum_{i=0}^{1}\sum_{j=0}^{2}\theta_{\frac{j}30}(3z|3\tau)\theta'_{\frac{i}2-\frac{j}3,0}(0|6\tau)\times$$
\eq{\times\left(6\theta_{00}(0|\tau)\theta_{0\frac{1}3}(0|\tau)\theta'_{0\frac{1}3}(0|\tau)
\theta'_{\frac{i}2\frac{2}3}(0|2\tau)+\theta_{00}(0|\tau)\theta_{0\frac{1}3}(0|\tau)\theta''_{0\frac{1}3}(0|\tau)
\theta_{\frac{i}2\frac{2}3}(0|2\tau)+\right.}
 $$+\left.
2\theta_{00}(0|\tau)\brc{\theta'_{0\frac{1}3}(0|\tau)}^2\theta_{\frac{i}2\frac{2}3}(0|2\tau)-
\theta''_{00}(0|\tau)\brc{\theta_{0\frac{1}3}(0|\tau)}^2\theta_{\frac{i}2\frac{2}3}(0|2\tau) \right)=0.$$
 Therefore,
(\ref{eq:consEl}) is equivalent to the following relation on theta constants:
  $${6\theta_{00}(0|\tau)\theta_{0\frac{1}3}(0|\tau)\theta'_{0\frac{1}3}(0|\tau)
\brc{\theta'_{\frac{2}30}(0|6\tau)\theta'_{0\frac{2}3}(0|2\tau)+
\theta'_{\frac{1}60}(0|6\tau)\theta'_{\frac12\frac{2}3}(0|2\tau)}+ }$$
  \eq{+\brc{\theta_{00}(0|\tau)\theta_{0\frac{1}3}(0|\tau)\theta''_{0\frac{1}3}(0|\tau)
+2\theta_{00}(0|\tau)\brc{\theta'_{0\frac{1}3}(0|\tau)}^2- \theta''_{00}(0|\tau)\brc{\theta_{0\frac{1}3}(0|\tau)}^2}\times }
  $${\times\brc{\theta'_{\frac{2}30}(0|6\tau)\theta_{0\frac{2}3}(0|2\tau)+
\theta'_{\frac{1}60}(0|6\tau)\theta_{\frac12\frac{2}3}(0|2\tau)}=0\,.}$$
 The latter equation can be separated into the other
two identities of simpler form:
 \eqlb{eq:sId1}{\theta_{00}(0|\tau)\theta_{0\frac{1}3}(0|\tau)\theta'_{0\frac{1}3}(0|\tau)=
-2\imath\sqrt{3}\, \theta_{\frac{1}30}(0|3\tau)\brc{\theta'_{\frac{2}30}(0|6\tau)\theta_{0\frac{2}3}(0|2\tau)+
\theta'_{\frac{1}60}(0|6\tau)\theta_{\frac12\frac{2}3}(0|2\tau)}, }
  \eq{\theta_{00}(0|\tau)\theta_{0\frac{1}3}(0|\tau)\theta''_{0\frac{1}3}(0|\tau)
+2\theta_{00}(0|\tau)\brc{\theta'_{0\frac{1}3}(0|\tau)}^2- \theta''_{00}(0|\tau)\brc{\theta_{0\frac{1}3}(0|\tau)}^2= }
  $${=12\imath\sqrt{3}\, \theta_{\frac{1}30}(0|3\tau)\brc{\theta'_{\frac{2}30}(0|6\tau)\theta'_{0\frac{2}3}(0|2\tau)+
\theta'_{\frac{1}60}(0|6\tau)\theta'_{\frac12\frac{2}3}(0|2\tau)}\,. }$$
 Analyzing the relation (\ref{eq:sId1}), we find it to
be the consequence of the following identity:
 \eq{\theta_{\frac16\frac12}\brc{0|\tau}\theta'_{\frac13
0}(0|\tau)=\frac{2\pi}3\, \theta^2_{\frac16\frac12}\brc{0|\frac{\tau}2}\theta^2_{\frac16\frac12}\brc{0|2\tau}\,.}
This is a
specialization of the Macdonald identity for the case of \eqs{BC_1} affine root system \cite{Mac'72}.

\section{Connection with the Ruijsenaars models}
\label{sec:Ruj} The integrable systems under consideration were  proposed in \cite{BMMM'2000,MM} as systems p,q-dual to the
elliptic Calogero and Ruijsenaars models. Thus, in some specific trigonometric limits they become dual to the trigonometric
Calogero and Ruijsenaars models and the Hamiltonians (\ref{eq:thetaHam}) become the Hamiltonians of the rational and
trigonometric Ruijsenaars models. The goal of this section is to describe these trigonometric limits and to derive the
limiting dependence of the coordinates $\textbf{z}$ and the elements of the period matrix $\Omega$ on the particle
coordinates and momenta \eqs{\brc{\textbf{u},\textbf{v}}} of the Ruijsenaars models.

At first, we recall the general construction  of the Hamiltonians for the $N$-particle integrable systems p,q-dual to the
elliptic Calogero and Ruijsenaars models. The generating function for these Hamiltonians is the genus-$N$ theta function on
the Jacobian of the Calogero spectral curve, where $N\times N$ period matrix $T(\textbf{a})$, depending on $N-1$ flat moduli
$\textbf{a}$, possesses the following property (reflecting the decoupling of the center-of-mass motion):
\eqlb{eq:Csc}{\sum_{j=1}^{N}T_{ij}(\textbf{a})=\tau,\quad \forall i,} where $\tau$ does not depend on $\textbf{a}$. Due to
(\ref{eq:Csc}) the genus-$N$ theta function decomposes into bilinear combinations of genus-one and genus-$g$ (\eqs{g=N-1})
theta functions. Then the claim of \cite{BMMM'2000,MM} is that all the ratios of these genus-$g$ theta functions are
Poisson-commuting with respect to the Seiberg-Witten symplectic structure.

In the case of three particles we have \eqs{3\times3} period   matrix $T$ with the property (\ref{eq:Csc}). Again, we do not
use any Seiberg-Witten data here. Introducing the notations for the genus-$2$ theta functions with characteristics
\eq{\theta^{(2)}\bsq{\textbf{b}}\equiv\theta^{(2)}\bsq{\at{\textbf{b}/3}{\textbf{0}}}=\sum_{\textbf{n}\in\mathbb{Z}^2}
\rme\brc{\dfrac12\brc{\textbf{n}+\textbf{b}/3}^t\,\Omega\,\brc{\textbf{n}+\textbf{b}/3}+
\brc{\textbf{n}+\textbf{b}/3}\cdot\textbf{z}},\quad \textbf{b}\in\mathbb{Z}^2,} we get the following decomposition of the
genus-$3$ theta function on the Jacobian of the Calogero spectral curve: \eq{\theta^{(3)}(\widetilde{\textbf{z}}|T)=
\theta_{00}(z_3|3\tau)\brc{\theta^{(2)}\bsq{00}+\theta^{(2)}\bsq{12}+\theta^{(2)}\bsq{21}}+ \theta_{\frac130}(z_3|3\tau)
\brc{\theta^{(2)}\bsq{11}+\theta^{(2)}\bsq{02}+\theta^{(2)}\bsq{20}} +} \eqn{+\theta_{\frac230}(z_3|3\tau)
\brc{\theta^{(2)}\bsq{22}+ \theta^{(2)}\bsq{01}+\theta^{(2)}\bsq{10}},} where all genus-$2$ theta functions are functions of
\eqs{(3z_1,\,3z_2|\,9\Omega)}, \eq{z_1=\frac{\wt z_1-2\wt z_2+\wt z_3}3,\qq z_2=\frac{\wt z_1+\wt z_2-2\wt z_3}3,\qq
z_3=\frac{\wt z_1+\wt z_2+\wt z_3}3,} and the genus-$2$ period matrix $\Omega$ can be expressed through the diagonal elements
\eqs{t_i\equiv T_{ii}} as follows: \eq{\Omega=\brc{\matr{cc}{
t_2-\frac{\tau}3&\frac12(t_1-t_2-t_3)+\frac{\tau}6\\
\frac12(t_1-t_2-t_3)+\frac{\tau}6&t_3-\frac{\tau}3 }}.} Then the
Poisson-commuting ratios are of the form \eqlb{eq:equivHam}{\wt
H_1=\dfrac{\theta^{(2)}\bsq{11}+
\theta^{(2)}\bsq{02}+\theta^{(2)}\bsq{20}}{\theta^{(2)}\bsq{00}+
\theta^{(2)}\bsq{12}+\theta^{(2)}\bsq{21}},\qq \wt
H_2=\dfrac{\theta^{(2)}\bsq{22}+
\theta^{(2)}\bsq{01}+\theta^{(2)}\bsq{10}}{\theta^{(2)}\bsq{00}+
\theta^{(2)}\bsq{12}+\theta^{(2)}\bsq{21}}.} The latter expressions
are rational functions of the Hamiltonians (\ref{eq:thetaHam}) due
to the particular case (\eqs{g=2,\,l=3}) of the following transition
formula:
\eq{\theta^{(g)}\bsq{\at{\textbf{0}}{\textbf{b}/l}}(\textbf{z}|l^{-1}\Omega)=
\sum_{0\leq c_i<l}
\rme\brc{\dfrac{\textbf{b}\cdot\textbf{c}}{l}}\theta^{(g)}\bsq{\at{\textbf{c}/l}{\textbf{0}}}(l\textbf{z}|l\Omega),\quad
0\leq b_i<l.} Thus, functions (\ref{eq:equivHam}) can be considered
as an equivalent pair of the Hamiltonians.

The described above projection method provides the  parameter $\tau$
of an elliptic curve, which can be associated with the elliptic
Calogero and Ruijsenaars models. So, it is natural to consider the
trigonometric limit \eqs{\textrm{Im}\,\tau\rightarrow+\infty}.
Additionally, we assume that there exists nonzero positive constant
\eqs{k>0} such that \eqlb{eq:spme}{t_i-\frac{\tau}3=\tau_i+k\tau,\qq
i=1,2,3} and all \eqs{\tau_i} have a finite nonzero limits:
\eq{\lim_{\textrm{Im}\,\tau\rightarrow+\infty}\tau_i=\wt\tau_i.}
Also, the coordinates \eqs{\widetilde{\textbf{z}}} on the Jacobian
of the Calogero spectral curve are the following functions of
particle momenta: \eq{\lim_{\textrm{Im}\,\tau\rightarrow+\infty}\wt
z_i=v_i,\qq i=1,2,3.} Then for the genus-$2$ theta function
\eqs{\theta^{(2)}\bsq{\textbf{b}}} we have the expansion
\eqn{\theta^{(2)}\bsq{b_1\,b_2}(3z_1,\,3z_2|\,9\Omega)=\sum_{m,n\in\mathbb{Z}}
q^{\frac{9}2k((m+b_1/3)^2+(n+b_2/3)^2-(m+b_1/3)(n+b_2/3))}\brc{\dfrac{q_1}{q_2q_3}}^{\frac92(m+b_1/3)(n+b_2/3)}\times}
\eq{\times q_2^{\frac92(m+b_1/3)^2}q_3^{\frac92(n+b_2/3)^2}
\rme((m+\frac{b_1}3)3z_1+(n+\frac{b_2}3)3z_2),} where
\eqs{q\equiv\rme(\tau)} and \eqs{q_i\equiv\rme(\tau_i)}. This gives
the  limits of the equivalent Hamiltonians (\ref{eq:equivHam}) in
the center of a mass frame:
\eqlb{eq:hLim1}{h_1=\lim_{\textrm{Im}\,\tau\rightarrow+\infty}q^{-\frac{k}{2}}\wt
H_1 = \rme(v_1)\rme\brc{\frac{\wt\tau_1}2}
+\rme(v_2)\rme\brc{\frac{\wt\tau_2}2}+
\rme(v_3)\rme\brc{\frac{\wt\tau_3}2},}
\eqlb{eq:hLim2}{h_2=\lim_{\textrm{Im}\,\tau\rightarrow+\infty}q^{-\frac{k}{2}}\wt
H_2= \rme(-v_1)\rme\brc{\frac{\wt\tau_1}2}
+\rme(-v_2)\rme\brc{\frac{\wt\tau_2}2}+
\rme(-v_3)\rme\brc{\frac{\wt\tau_3}2}.}

In the case of the system dual to the elliptic Calogero  model we obtain the following dependence on the particle coordinates
\eqs{\textbf{u}}: \eq{\rme\brc{\frac{\wt\tau_i}2}=\prod_{j\neq i}\sqrt{1-\frac{g^2}{(u_i-u_j)^2}},} and functions
(\ref{eq:hLim1}), (\ref{eq:hLim2}) are the Hamiltonians of the rational Ruijsenaars model.

In the case of the system dual to the elliptic Ruijsenaars model we obtain
\eq{\rme\brc{\frac{\wt\tau_i}2}=\prod_{j\neq i}\sqrt{1-\frac{g^2}{\sinh^2\brc{u_i-u_j}}}}
and functions (\ref{eq:hLim1}), (\ref{eq:hLim2}) are the Hamiltonians of the trigonometric Ruijsenaars model.

\section{Lax pair for $N=2$}
\label{sec:Lax} In this section we propose $2\times2$ Lax pairs for two-particle dual to elliptic models including dual to
the Double Elliptic one.
In general, using the original Ruijsenaars' idea \cite{Ruijs} (see also \cite{Feher}) one can find the Lax pair for the dual
model in the following way. Let an initial model is given by its $N\!\times\!N$ Lax matrix $L$ and the problem is to find its
dual Lax formulation ($\tilde L$). Taking into account group-theoretical interpretation of the duality we can formulate it in
terms of the linear (eigenvalue) problems
 \begin{equation}
  \label{p1}\begin{array}{c} L(p,q)\Psi=\Psi  \lambda (U)\,,\ \ U=\hbox{diag}(u_1,...,u_N)\,,\\ \ \\
 {\tilde L(v,u)}{\Psi^{-1}}=\Psi^{-1} f(Q)\,, \ \ Q=\hbox{diag}(q_1,...,q_N)\,,\end{array}
 \end{equation}
where $p,q$ -- canonical variables of the initial model, while $v,u$ are those of the dual one and $\Psi$ is the matrix of
eigenvectors.
 From (\ref{p1}) it follows that
  \begin{equation}
  \label{p2} \hbox{tr} (\la^k(U){\tilde L}^m)=\hbox{tr}\left(L^k f^m(Q)\right)
  \end{equation}
%
Functions $\la$ and $f$ define the free Hamiltonians\footnote{We assume that there exists a coupling constant $\nu$. The free
Hamiltonians appear at $\nu=0$. In the examples below we also use  notation $\nu'=\sqrt{-2}\nu$.} because $\tr L^k$ or $\tr
\ti L^k$ are symmetric functions of $\la(u_i)$ or $f(q_i)$ correspondingly. The free hamiltonians should satisfy the
anticanonicity conditions \cite{BMMM'2000}.

The idea of our construction is to use the anticanonicity conditions
together with (\ref{p1}). Let us demonstrate the derivation of the
dual Lax matrix in the simplest case $N\!=\!2$. In this case
$Q=\hbox{diag}(q,-q)$, $U=\hbox{diag}(u,-u)$ and there is  a single
Hamiltonian for each model (and there are no any non-trivial
identities discussed above). The Hamiltonians $H(p,q,\nu)$ (its free
version is $H_0(p)=H(p,q,0)$) and $\ti H(v,u,\nu)$ ($\ti H_0(v)$)
satisfy the following relations:
 \begin{equation}
  \label{qr002}
  H(p,q)= H_0(u)\,,\ \ \ti H(v,u)=\ti H_0(q)
 \end{equation}
while the anticanonicity condition $\{H(p,q),\ti H_0(q)\}_{p,q}=\{H_0(u),\ti H(v,u) \}_{u,v}$ takes the form:
  \begin{equation}
  \label{qr004}
  \p_p H(p,q)  \,\p_q\ti H_0(q) = \p_u H_0(u)\,\p_v \ti H(v,u)\,.
 \end{equation}
Next, we find $\Psi$ from known $L$. Some freedom in the definition of $\Psi$ can be fixed by condition
$\Psi\!\left.\right|_{\nu\!=\!0}=1$. We have
   \begin{equation}
  \label{qr013}
 \Psi=\mat{1}{\frac{\la_2-L_{22}}{L_{21}}}{\frac{\la_1-L_{11}}{L_{12}}}{1}\,,\ \ \
 \det\Psi=\frac{1}{L_{12}L_{21}}\left(\la_1(L_{22}\!-\!\la_2)+\la_2(L_{11}\!-\!\la_1)\right)\,,
 \end{equation}
where $\la_1=\la(u)$, $\la_2=\la(-u)$ and, therefore,
   \begin{equation}
  \label{qr014}
  \begin{array}{l}
\ti L=\frac{1}{\la_1(L_{22}\!-\!\la_2)\!+\!\la_2(L_{11}\!-\!\la_1)}
\mat{f_1L_{12}L_{21}\!-\!f_2(\la_1\!-\!L_{11})(\la_2 \!-\!L_{22}) }{ L_{12}(\la_2-L_{22})(f_1-f_2) }
{-L_{21}(\la_1-L_{11})(f_1-f_2)}{f_2 L_{12}L_{21}\!-\!f_1(\la_1\!-\!L_{11})(\la_2 \!-\!L_{22})}\,,
 \end{array}
 \end{equation}
where $f_1=f(q)$, $f_2=f(-q)$.


\paragraph{Remark} Using (\ref{qr014}) together with (\ref{p2}) for $k\!=\!m\!=\!1$  and (\ref{qr004}) one
can easily reproduce the "trigonometric Calogero -
rational Ruijsenaars" models duality \cite{Ruijs} for $N=2$ via $\la_{1,2}(u)=\pm u$ and $f_{1,2}=\exp(\pm q\eta)$:
 \begin{equation}
  \label{qr020}
 L=\mat{p}{\frac{\nu'\eta}{\sinh(q\eta)}}{-\frac{\nu'\eta}{\sinh(q\eta)}}{-p}\ \ \ \longrightarrow\ \ \ \
 \ti L=\mat{e^{\eta v}\sqrt{1-\frac{\nu'^2\eta^2}{u^2}}}{-\frac{\nu'\eta}{u}}
  {\frac{\nu'\eta}{u}}{e^{-\eta v}\sqrt{1-\frac{\nu'^2\eta^2}{u^2}}}
 \end{equation}
 where $\eta$ is the inverse "speed of light".

\subsection*{Elliptic Calogero Model and its Dual}
The elliptic two-particle Calogero model is defined by
  \begin{equation}
  \label{p3}{
L=\mat{p}{\frac{\nu'}{\sn(q)}}{-\frac{\nu'}{\sn(q)}}{-p},\ \ \
 -\frac{1}{2}\det L=H(p,q)=\frac{1}{2}\left(p^2-\frac{\nu'^2}{\sn^2(q)}\right)
 =\frac{p^2}{2}+\frac{\nu^2}{\sn^2(q)}\,, }
  \end{equation}
where $\nu'=\sqrt{-2}\nu$. With $\la(u)=u$ the Hamiltonian $H(p,q)\!=\!H_0(u)\!=\!\frac12 u^2$ and the matrix (\ref{qr014})
is simplified:
 \begin{equation}
  \label{qr031}
\ti L=\mat{\frac{1}{2}(f_1+f_2)+\frac{L_{11}}{2u}(f_1-f_2)}{-\frac{1}{2u}L_{12}(f_1-f_2)}
{-\frac{1}{2u}L_{21}(f_1-f_2)}{\frac{1}{2}(f_1+f_2)-\frac{L_{11}}{2u}(f_1-f_2)},
 \end{equation}
To determine the function $f$  set $\ti H=\ti H(v,u)=\tr \ti L=\cn(q)$. This gives
 \begin{equation}
  \label{qr032}
 f_1+f_2=\cn(q)
 \end{equation}
Next, let us use the condition (\ref{p2}) $\tr (\la(U)\ti L)=\tr\left(L f(Q)\right)$:
  \begin{equation}
  \label{qr033}
u(\ti L_{11}- \ti L_{22})=p(f_1-f_2)\,.
 \end{equation}
Together with the anticanonicity condition $p\ \cn'(q)=u\frac{\p \ti H(v,u)}{\p v}$ it is natural to set
  \begin{equation}
  \label{qr034}
f_1-f_2=\cn'(q)=-\sn(q)\dn(q),\ \ \ti L_{11}- \ti L_{22}=\p_v \ti H(v,u)=\dot{u}
 \end{equation}
Thus,
  \begin{equation}
  \label{p31}
  \la(u)=u\,,\ \ \ f(q)=\frac{1}{2}(\cn(q)+\cn'(q))=\frac{1}{2}(\cn(q)-\sn(q)\dn(q))\,.
    \end{equation}
Plugging into (\ref{qr014}) we get the dual Lax matrix
  \begin{equation}
  \label{p4}{ \ti L=\frac{1}{2}\mat{ \cn(q)+\dot{u} }{ -\dn(q)\frac{\nu'}{u} }{\dn(q)\frac{\nu'}{u}}{\cn(q)-\dot{u}}\,,
\ \ \dot{u}=\p_v \ti H(v,u)\,. }
 \end{equation}
Notice that the equation
  \begin{equation}
  \label{qr027}
  \det \ti L=f(q)f(-q)\,,
 \end{equation}
which follows from (\ref{p1}) is not an identity  since we have already used the anticanonicity condition (it becomes the
identity if to express back $\dot u$). Instead, it provides the equation for the dual Hamiltonian:
  \begin{equation}
  \label{qr040}
\left({\p_v \ti H}\right)^2=\left(\alpha^2(u)-\ti H^2\right)(\kappa'^2+\kappa^2 \ti H^2)\,,\ \ \
\alpha^2(u)=1-2\frac{\nu^2}{u^2}\,,
 \end{equation}
 where $\kappa$ and $\kappa'$ are the standard elliptic moduli. Then we get the known answer \cite{BMMM'2000}:
  \begin{equation}
  \label{qr041}
H(v,u)=\alpha(u) \cn\left(v\beta(u)\Big|\frac{\kappa\alpha(u)}{\beta(u)}\right)\,,\ \ \
\beta(u)=\sqrt{\kappa'^2+\kappa^2\alpha^2(u)}\,.
 \end{equation}
The dual moduli are $\ti\kappa=\frac{\kappa\alpha(u)}{\be(u)}$ and $\ti\kappa'=\frac{\kappa'}{\beta(u)}$. It is easy to
verify that $\dn(q|\kappa)=\beta(u)\dn(\beta(u)v|\ti\kappa)$. Then the Lax matrix (\ref{p4}) takes the form:
  \begin{equation}
  \label{p455}
  \ti L=
  \frac{1}{2}\mat{ \alpha\left(\cn(\beta v|\ti\kappa)-\beta\sn(\beta v|\ti\kappa)\dn(\beta v|\ti\kappa)\right) }
  { -\beta\dn(\beta v|\ti\kappa)\frac{\nu'}{u} }{\beta\dn(\beta v|\ti\kappa)\frac{\nu'}{u}}
  {\alpha\left(\cn(\beta v|\ti\kappa)+\beta\sn(\beta v|\ti\kappa)\dn(\beta v|\ti\kappa)\right) }
 \end{equation}
Being written in the form (\ref{p4}) (in terms of $\dot u,
u$-variables) the dual Lax matrix is similar to the one of the
rational Calogero model. Then it is easy  to find its M-operator
(since $\cn(q)$ and $\dn(q)$ are constants on the dual Hamiltonian
flow)
  \begin{equation}
  \label{p404}
   \ti M=\mat{0}{-\dn(q)\frac{\nu'}{2u^2}}{-\dn(q)\frac{\nu'}{2u^2}}{0} =
   \mat{0}{-\beta\dn(\beta v|\ti\kappa)\frac{\nu'}{2u^2}}{-\beta\dn(\beta v|\ti\kappa))\frac{\nu'}{2u^2}}{0}\,,
   \end{equation}
which provides equation of motion from the Lax equation $\dot{\ti L}=[\ti L,\ti M]$:
  \begin{equation}
  \label{qr038}
\ddot{u}=-\frac{\nu'^2}{u^3}\dn^2(q)=2\frac{\nu^2}{u^3}\dn^2(q)\,.
 \end{equation}
%

\subsection*{Double Elliptic Model}
In this case the Lax pair can  be constructed by a simple
generalization of the previously obtained (\ref{p455}),
(\ref{p404}):
  \begin{equation}
  \label{p4553}
  \ti L=
  \frac{1}{2}\mat{ \alpha\left(\cn(\beta v|\ti\kappa)-\beta\sn(\beta v|\ti\kappa)\dn(\beta v|\ti\kappa)\right) }
  { -\beta\dn(\beta v|\ti\kappa)\vf(u) }{\beta\dn(\beta v|\ti\kappa)\vf(u)}
  {\alpha\left(\cn(\beta v|\ti\kappa)+\beta\sn(\beta v|\ti\kappa)\dn(\beta v|\ti\kappa)\right) }\,,
 \end{equation}
  \begin{equation}
  \label{p4043}
   \ti M=
   \mat{0}{\frac12\beta\dn(\beta v|\ti\kappa)\vf'_u(u)}{\frac12\beta\dn(\beta v|\ti\kappa))\vf'_u(u)}{0}\,,
   \end{equation}
where $\alpha=\alpha(u)=\sqrt{1+\frac{\nu'^2}{\sn^2(u)}}$ while the
function $\vf(u)$ can be chosen to be $\vf(u)=\frac{\nu'}{\sn(u)}$.
Another  possible  choice is $\vf(u)=\alpha(u)$.

We hope to generalize the suggested approach for $N>2$ in our future
papers. The final purpose is to get the Lax pair with spectral
parameter. As is well known the elliptic many-body systems are gauge
equivalent to the top-like models \cite{LOZ}. The later are
described by quadratic algebras \cite{CLOZ}. Therefore, another
possibility is to find a generalizations of the higher rank Sklyanin
algebras and corresponding r-matrices. Some progress towards this
idea was made in \cite{BGOR}. One more approach for constructing the
Lax pairs $L(z)$ is based on the modification of bundles $\Xi(z)$
via $L(z)=\Xi^{-1}(z)f(\p_z)\Xi(z)$, where $f$ is some function
(which can be elliptic) \cite{AASZ}.

\section{Conclusion}

We found explicit theta-function   identities of genus two that
stand behind the Poisson commutativity of the Hamiltonians
(\ref{eq:thetaHam}) of the three-particle integrable systems with
elliptic dependence on momenta. The problem of defining the elements
of the Poisson structure \eqs{P_{ijk}} was reduced to the problem of
solving the system of three partial differential equations with
respect to the pair of unknown functions. In addition, we
constructed Lax representations for the two-particle systems, which
still need to be extended to many-particle case.

An open problem is to find  the set  of all solutions of the system
of three partial differential equations (\ref{eq:detPS0}) and to
understand whether the three-particle integrable systems under
consideration are unique or not. Solutions of the system
(\ref{eq:detPS0}) define the Poisson structure in terms of the
coordinates on the Jacobian of the spectral curve and the elements
of its period matrix. To establish the connection with the
Seiberg-Witten symplectic structure one should transform the Poisson
brackets into the canonical ones. This coordinate transformation
acts only on the elements of the period matrix and gives an explicit
dependence of the period matrix on the flat moduli of the
corresponding spectral curve. In this way we can get some useful
information about the algebraic geometry of the spectral curves. We
are going to cover these topics in the subsequent papers.

 {\normalsize{\bf Acknowledgments.}} We are grateful  to
 S. Arthamonov, A. Levin and M. Olshanetsky for helpful discussions.
 The work was partly supported by
 Ministry of Education and Science of the Russian Federation
under contract 8207 (A.Mir., A.Mor.,  A.Z.), the Brazil National
Counsel of Scientific and Technological Development (A.Mor.), by
NSh-3349.2012.2 (A.Mir., A.Mor.), by RFBR grants 13-02-00457
(A.Mir.), 13-02-00478 (A.Mor.) and 12-01-00482 (G.A., A.Z.), by
joint grants 12-02-92108-Yaf (A.Mir., A.Mor.), 13-02-91371-ST
(A.Mir., A.Mor.), 14-01-93004-Viet (A.Mir., A.Mor.), by leading
young scientific groups RFBR 12-01-33071 mol$\_$a$\_$ved (G.A.,
A.Z.) and by D. Zimin's fund "Dynasty" (A.Z.).



\footnotesize{

}
\end{document}